\documentclass[prl,twocolumn,superscriptaddress,showpacs]{revtex4}
\usepackage{graphicx}
\usepackage{epsfig}
\usepackage{amssymb,amsmath,amsfonts,hyperref}
\usepackage{wasysym}
\usepackage{latexsym}

\newcommand{\be}{\begin{eqnarray}}
\newcommand{\ee}{\end{eqnarray}}

\begin{document}
\title{Fractional spin textures in the frustrated magnet  SrCr$_{9p}$Ga$_{12-9p}$O$_{19}$}
\author{Arnab Sen}
\affiliation{\small{Tata Institute of Fundamental Research, 1, Homi Bhabha Road, Mumbai
400005, India}}
\affiliation{\small{Department of Physics, Boston University, 590 Commonwealth Avenue, Boston, Massachusetts 02215, USA}}
\author{Kedar Damle}
\affiliation{\small{Tata Institute of Fundamental Research, 1, Homi Bhabha Road, Mumbai
400005, India}}
\author{Roderich Moessner}
\affiliation{\small{Max-Planck-Institut f\"{u}r Physik komplexer Systeme, 01187 Dresden, Germany }}

\begin{abstract}
{We consider the archetypal 
frustrated antiferromagnet  SrCr$_{9p}$Ga$_{12-9p}$O$_{19}$ in its well-known spin-liquid state, and demonstrate that a
Cr$^{3+}$ spin $S=3/2$ ion in direct proximity to a pair of vacancies (in disordered $p<1$ samples)  is cloaked by a spatially extended spin texture that encodes the correlations of the parent spin-liquid. In this
spin-liquid regime, the combined object
has a magnetic response identical to a classical spin of length $S/2 = 3/4$, which dominates over the small
intrinsic susceptibility of the pure system. This fractional-spin texture leaves an unmistakable imprint on the measured $^{71}$Ga nuclear magnetic resonance (NMR) lineshapes, which we compute using Monte-Carlo simulations and compare
with experimental data.}

\end{abstract}

\pacs{75.10.Jm 05.30.Jp 71.27.+a}
\vskip2pc

\maketitle

\begin{figure}
{\includegraphics[width=\hsize]{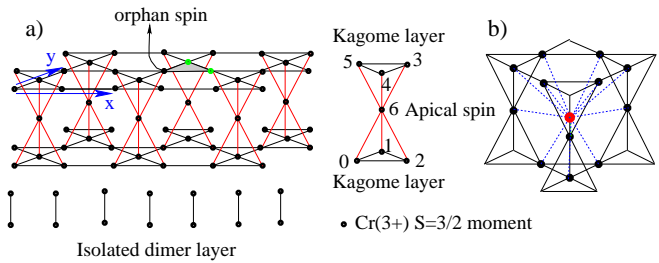}}
\caption{ a) The Cr atoms in
SCGO form a lattice made up of kagome bilayers separated from each other
by a layer of isolated dimers consisting of pairs of Cr atoms. Each kagome
bilayer is a corner sharing arrangement of tetrahedra and triangles made up 
of two Kagome lattices that are coupled to each other through ``apical''
Cr sites shared between up-pointing and down-pointing tetrahedra.
Links between near-neighbour Cr sites in each bilayer represent a Heisenberg
exchange coupling $J = 80$ K between neighbouring Cr$^{3+}$ spins, while
links in the isolated dimer layer represent a Heisenberg exchange coupling
$J' = 216$ K between the two Cr$^{3+}$ ions that constitute each pair. Two
vacancies in a triangle (green circles) leave behind an orphan spin. b)
The $12$ Cr$^{3+}$ sites (black) that are hyperfine coupled to a given Ga(4f)
site (red) in the SCGO lattice.}
\label{Fig1_lattice}
\end{figure}

SrCr$_{9p}$Ga$_{12-9p}$O$_{19}$ (SCGO) is a remarkable magnetic material which does not show any signs
of magnetic ordering even at extremely low temperatures $T \sim \Theta_{CW}/100$, where
$\Theta_{CW} \approx 500$ K is the so-called Curie-Weiss temperature at which mean-field theory predicts magnetic ordering of its corner-sharing network (Fig~\ref{Fig1_lattice}) of antiferromagnetically coupled Cr$^{3+}$ $S=3/2$ moments.
The original observation\cite{ober1,ober2,ramirez} of this broad regime of spin-liquid behaviour in SCGO
led to an intensive study of a series of samples with
varying density $x \equiv 1-p$ of vacancies in the Cr$^{3+}$ spin network, including NMR\cite{Limot_etal_prb}, susceptibility\cite{schifferSUS}, and $\mu$SR\cite{kerenMUSR} experiments. Thus, SCGO is among the best known and most systematically studied candidates for spin-liquid behaviour in frustrated magnets. 

One key observation, due to Schiffer and Daruka, was the presence of a paramagnetic ``Curie tail'' in the low temperature uniform susceptibility of these compounds, 
well modeled by a defect contribution $\chi_{\mathrm{defect}} = C_d/T$ that
dominates over the intrinsic susceptibility $\chi_{\mathrm{intrinsic}} \approx C_1/(T+\Theta_{CW})$ for $T \ll \Theta_{CW}$; the Curie constant $C_d$ in this ``two-population''
phenomenology was associated with a population of  paramagnetic objects dubbed ``orphan spins''
\cite{schifferSUS}. Another important observation, due originally to Limot~{\em et. al.}\cite{Limot_etal_prb}, is the
{\em apparently symmetric} NMR line broadening $\Delta H$ that scales as $\Delta H \propto x/T$ for not-too-low
$x$ and $T$; this was in turn interpreted phenomenologically as a signature of a disorder-induced spin-texture---a short-ranged oscillating spin density profile induced by some lattice defects. 

On the theoretical side, Berlinksy and one of us used a ``single-unit approximation''\cite{Moessner_berlinsky} to predict that ``defective'' simplices (corner-sharing units), in which all but one spin has been 
substituted for with non-magnetic impurities, must give rise to Curie tails
in the low temperature susceptibility of isotropic classical spin-$S$ antiferromagnets
on corner-sharing lattices such as SCGO. In related work, Henley used classical energy
minimization considerations at $T=0$ to argue that an infinitesimal magnetic field applied to
such a system with a single defective simplex should induce a spin-texture with saturation magnetization $S/2$\cite{Henley_2000}.
This  suggested that the phenomenological orphan spins\cite{schifferSUS} and oscillating spin textures\cite{Limot_etal_prb} invoked earlier are related to the presence of such defective simplices, with the lone spins
on such defective simplices providing a microscopic basis for the phenomenological orphan spin population
of Ref~\onlinecite{schifferSUS}.
Although this orphan-texture complex (comprising the orphan spin on a defective simplex and its surrounding
spin texture) has thus been implicated in some of the most intriguing experimental observations
on SCGO, a fundamental understanding of it has been lacking.

Here, we develop a quantitatively accurate analytical theory that provides a full characterization of
this orphan-texture complex by accounting for both entropic and energetic effects in the spin-liquid regime of low temperatures ($\bar{T} \equiv k_BT/JS^2 \ll 1$) and low magnetic field $h$ ($\bar{h} \equiv g_L\mu_Bh/JS \ll 1$), but
arbitrary $\bar{h}/\bar{T}$.  
In this regime, we find that the orphan spin magnetisation is equal to the magnetisation of {\em a spin $S$ in a field $h/2$}. This
is accompanied by an extended spin-texture, which is scale-free at $T=0$ but acquires a finite extent as $T$ is increased. We determine the intricate pattern of spin correlations in the texture and demonstrate that its net magnetisation {\em cancels off half of the orphan spin's moment}, thus giving rise to an orphan-texture complex that behaves as a {\em classical spin $S/2$ in field $h$}, with a susceptibility given by 
\begin{displaymath}
\chi_{S/2}(T) = (g_L\mu_BS/2)^2/3k_BT
\end{displaymath}
This provides a particularly dramatic instance of fractionalisation
of spins in a simple classical system.

In remarkable correspondence with experiment\cite{Limot_etal_prb}, we find that
these extended, fractional-spin orphan-texture complexes
show up prominently in SCGO at not-too-low $x=1-p$ as a large, nearly symmetric low temperature broadening $\Delta H \propto x/T $ of our predictions for the $^{71}$Ga(4f) NMR line\cite{Limot_etal_prb} associated with Ga nuclei at the so-called
(4f) crystallographic position in the SCGO lattice (Fig~\ref{Fig1_lattice}).
We therefore
focus below
on this particular case, although our analytical low temperature, low field
results apply more generally to orphan-texture complexes surrounding
isolated defective simplices
in kagome, pyrochlore and SCGO lattices.

SCGO is described by the classical Hamiltonian:
\be
 {\cal H} = \frac{J}{2}\sum_{\XBox} (\sum_{i \in \XBox}\vec{S}_i - \frac{g_L\mu_B\vec{h}}{2J})^2 + \frac{J}{2}\sum_{\triangle} (\sum_{i \in \triangle}\vec{S}_i - \frac{g_L\mu_B\vec{h}}{2J})^2 \nonumber
\label{eq2}
\ee
where $\vec{S}_i$ are classical length-$S$ spins ($S=3/2$ for SCGO) with
saturation moment $g_LS$ in units of the Bohr magneton $\mu_B$ ($g_L=2$ for
SCGO), $J$ is the nearest-neighbour Heisenberg exchange coupling ($J\simeq80 K$
for SCGO\cite{Limot_etal_prb}), and $\vec{h}$ is the external field. $\XBox$
refers to the tetrahedra that consisting of a triangle in a kagome layer
attached to an apical spin in the triangular layer (Fig~\ref{Fig1_lattice}),
while $\Delta$ refers to those triangles in the kagome layer which are not
associated with an apical spin.\begin{figure*}[!]
 {\includegraphics[width=2.1\columnwidth]{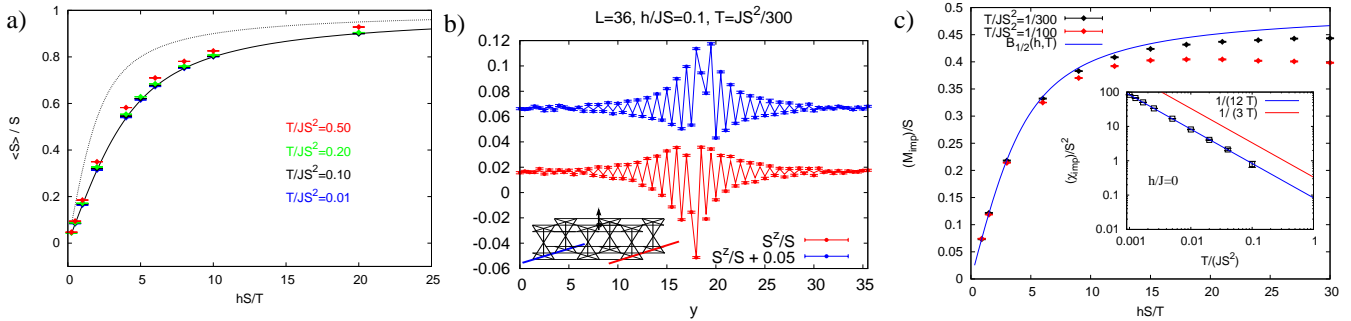}}
\caption{ a) The orphan spin in
  field $h$ develops magnetization (symbols) that matches that of an isolated
  spin $S$ in field $h/2$ (full line) in the spin liquid regime. Also shown
  (dotted curve) is the magnetization expected for an isolated spin $S$ in
  field $h$. b)  The orphan induced
  spin texture is shown along two cuts on the lattice. Numerical data is shown
  as points with statistical error bars and the effective theory result as
  lines in the main plot. The spin texture shown in red (blue) in main plot is
  along the cut shown in red (blue) in the inset. The orphan spin is denoted by
  an arrow in the inset. c) The impurity
  magnetization (defined as the difference in magnetic response between the
  diluted and the pure system) of a single orphan-texture
  complex (symbols) compared with $B_{S/2} (h,T)$ (solid curve), the asymptotic
  prediction deep in the spin-liquid regime. $B_{S/2}(h,T)$ is the
  magnetization curve of a classical spin $S/2$. Note the much faster
  convergence of numerical data to asymptotic predictions in the linear (small
  $hS/T$) part of the magnetization curve as compared with the non-linear
  regime at large $hS/T$. Inset: The corresponding impurity susceptibility
  (defined in text) matches the susceptibility of a free classical spin $S/2$ 
up to remarkably high temperatures $T \sim 0.1JS^2$. All Monte-Carlo data
exhibited in a) was extrapolated to the thermodynamic limit 
using a sequence of sizes up to $L=36$ (with $7L^2$ sites),
while that in c) was extrapolated using a sequence of sizes up to $L=20$.} 
\label{Fig2_omnibus}
\end{figure*}

Here, we model the pure $p=1$ system in terms of an effective free energy
functional ${\mathcal F}$ that incorporates the energetics of the
antiferromagnetic interactions $J$ on an equal footing with entropic effects of
thermal fluctuations:
\begin{displaymath}
{\mathcal F} =
{\mathcal H}(\{\vec{\phi}_i\})+\frac{T}{2}\sum_{i} \rho_i \vec{\phi}_i^2
\end{displaymath}
The unconstrained effective field $\vec{\phi}_i$ serves as a surrogate for the microscopic fixed-length spin variables $\vec{S}_i$, with
the statistical weight of a field configuration $\{\vec{\phi}_i\}$ being
proportional to $\exp(-{\mathcal F}/T)$. ${\mathcal H}(\{\vec{\phi}_i\})$ is the classical Hamiltonian of the system now
written in terms of $\vec{\phi}_i$, and the phenomenological stiffness constants $\rho_i$ are fixed by requiring that the
mean length of $\vec{\phi}_i$ equal $S$ within the effective theory: $\langle \vec{\phi}_i^2 \rangle_{\mathcal F} = S^2$\cite{sup}.
For a background to this approach for pure systems, see the review by Henley\cite{Henley_effectivetheory}.

To incorporate disorder effects in diluted $p < 1$ samples, we assume that the
stiffness constants 
do not change significantly from their pure values, but extend the effective
theory in two important ways: First, we model vacancies in the lattice by
setting
$\vec{\phi}_i$ to zero on vacancy sites. Secondly and more crucially, as the fixed length
nature of an orphan spin on a defective simplex is expected to play a central
role, we retain it as a microscopic length-$S$ spin and do not 
introduce an effective field variable at such sites.

We now use this effective theory to analyze the SCGO magnet with a single
defective triangle, {\em i.e.} 
two vacancies on one triangle of the SCGO lattice
(Fig~\ref{Fig1_lattice}). In this case, the effective theory
reduces to a length-S orphan spin $S\vec{n}$ ($\vec{n}^2=1$) coupled
to a constrained Gaussian theory for $\vec{\phi}_i$ (with constraints
$\vec{\phi}_i = 0$ at the two vacancy sites). Focusing first 
on the orphan spin $S\vec{n}$ in this defective triangle by integrating
out the $\vec{\phi}$ fields, we find\cite{sup} that the exchange
field from the surrounding spin liquid ``screens'' exactly half the external magnetic field on the orphan spin in the low temperature, low field
spin liquid regime, yielding a spin $S$ variable 
that ``sees'' a magnetic field $h/2$, and therefore develops a polarization
equal to that of a free classical spin $S$ in a field $h/2$. This striking prediction is fully confirmed by Monte-Carlo studies\cite{sup} of the classical SCGO
magnet with one defective triangle (see Fig~\ref{Fig2_omnibus} (a)), which also reveal that this prediction is surprisingly robust, remaining accurate for temperatures as high as $T \sim 0.1 JS^2$ for this example.

Next, we turn to a detailed description of the extended spin texture
that cloaks this orphan spin in the spin-liquid regime. In addition
to the uniform external field $\vec{h} = h\hat{z}$  that acts on all the $\vec{\phi}_i$ in the constrained Gaussian
action, this orphan spin polarization also gives rise to a local exchange field
that acts in the $\hat{z}$ direction on $\vec{\phi}_i$ at the two undiluted sites
adjacent to the orphan spin.
The extended spin texture surrounding the orphan spin is modeled within the effective theory
by calculating the response $\langle \phi^z_{i} \rangle_{\mathcal F}$
to these fields\cite{sup}. At $T = 0$,  the computed texture's envelope decays as $1/|\vec{r}|$ away from the orphan spin, while the overall scale is set by the orphan spin's saturation magnetization. At finite-$T$ and small fields in the spin-liquid regime, the power law envelope is cut off by a thermally introduced finite correlation length $\xi\sim1/ \sqrt{T}$, endowing it
with an effective size $\xi^d \sim 1/T^{d/2}$ in $d$ dimensions; in our
$d=2$ example, this gives a spatial size $\sim 1/T$ scaling identically
with the
overall magnetization scale of the texture, which is set by the orphan spin's susceptibility $\sim 1/T$. These predictions are compared with
Monte-Carlo simulation\cite{sup} results in Fig~\ref{Fig2_omnibus} (b)
for the case of a single defective triangle in SCGO, and the
agreement is remarkably good.

Although the computed texture decays slowly (as $1/|\vec{r}|$ at $T=0$), its oscillations conspire to reduce its net moment so that it {\em cancels out precisely half} of the orphan spin's moment, endowing the orphan-texture complex with a net ``impurity magnetization'' (excess magnetization over and above that of a pure system in the same external field) equal
to the magnetization of a classical spin $S/2$ in field $h$. 
The corresponding ``impurity susceptibility'' $\chi_{\mathrm{imp}}$ arising from
a single orphan-texture complex is $(g_L\mu_BS/2)^2/3k_BT$---a Curie tail dominating the low-$T$ response. 
For the case of a single defective triangle in SCGO, our Monte Carlo simulations validate this striking prediction up to surprisingly high temperatures of order $T \sim 0.1JS^2$ (Fig~\ref{Fig2_omnibus} (c)).

Fortunately, this novel physics leaves its imprint on the Ga(4f) NMR line, which is a sensitive probe of the distribution
of the twelve Cr spin polarizations that are hyperfine coupled to each Ga(4f) nucleus in SCGO (Fig~\ref{Fig1_lattice} (b)).
We calculate this distribution in diluted SCGO lattices
with uncorrelated site dilution probability $x=1-p$\cite{sup}, using the
experimentally known values\cite{Limot_etal_prb} of these hyperfine couplings, and Monte-Carlo simulations\cite{sup} to fully account for the physics of a non-zero density of orphan spins in order to obtain quantitatively accurate results. \begin{figure*}[!]
 {\includegraphics[width=2.1\columnwidth]{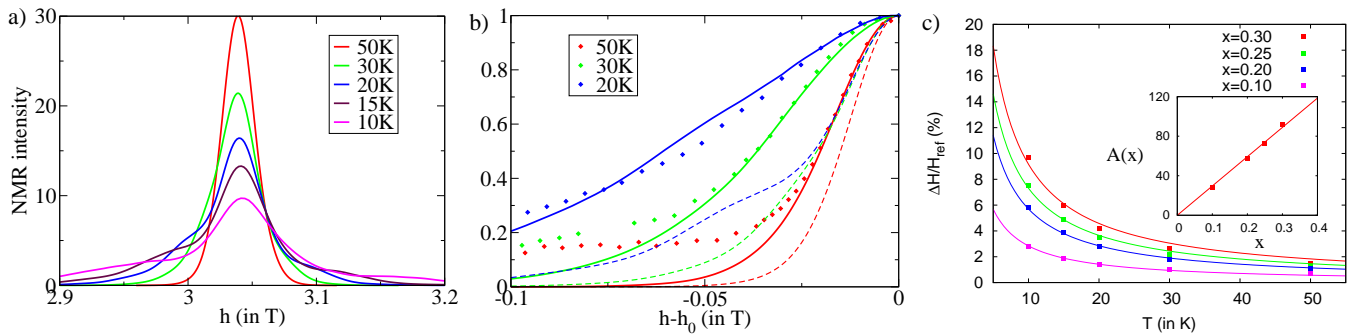}}
\caption{Predictions for Ga(4f) NMR lines. a) $^{71}$Ga(4f) ``field-scan'' NMR
  lines predicted within the minimal disorder model of uncorrelated dilution at
  vacancy 
  density $x=0.2$. (b) Experimental
results of Ref~{\protect{\cite{Limot_etal_prb}}} for field-scan $^{71}$Ga(4f)
NMR lines 
(dots) at dilution $x=0.19$, plotted using datafiles provided by P.~Mendels,
and compared to theoretical predictions of the minimal disorder model (with $y$
axis rescaled and $x$ axis offset appropriately for easy comparison of
linewidths and shapes of all curves) at vacancy density $x=0.2$ (dashed curves)
and $x=0.3$ (full curves). Only the low-field side of the experimental line is
compared with theory, as 
the high-field side of the experimental line is known to be contaminated by
the presence of a satellite peak arising from non-stoichiometric (impurity)
$^{71}$Ga nuclei located at three different Cr sites in the SCGO
lattice\cite{Limot_etal_prb}. (c) Temperature and $x$ dependence of the width $\Delta H$
  of the theoretically predicted field-scan $^{71}$Ga(4f) NMR lines, normalized
  by a reference field $H_{ref}=3.12$T. Solid curves in the  
main plot are fits of $\Delta H$
to ${\cal A}(x)/T$ over the experimentally relevant temperature range, and
inset shows the approximately linear $x$ dependence of the best-fit values of
${\cal A}(x)$ over the experimentally relevant range of $x$. The Monte-Carlo simulations were performed
using a sequence of sizes ranging from $L=16$ to $L=50$ (with $7L^2$ sites)
to eliminate finite-size effects, with on average $20$ disorder realizations
at each size.}
\label{Fig5}
\end{figure*}

As is clear from
the example at $x=0.2$ shown in Fig~\ref{Fig5} (a), this calculation
predicts a 
line that is broad and appears symmetric, reflecting
the fact that the spin textures are staggered and involve a very large
number of spins, making it difficult to discern the $O(1)$ net
moment of each texture that endows the line with a slight bias
towards lower magnetic fields.

In Fig~\ref{Fig5} (c), we show the
temperature and $x$ dependence of the width of our predicted lines  for not-too-small
values of $x$ and $T$, of greatest relevance to experiments. As the lineshape is not well-approximated
by a Gaussian, we do not fit the line to a Gaussian, but rather
use the definition $\Delta H \equiv 2 \sqrt{2\ln(2)}\left(\langle H^2 \rangle - \langle H \rangle^2\right)^{1/2}$ which
reduces to the standard value for a Gaussian line but provides an unbiased measure of the width in
more general cases. As is clear from Fig~\ref{Fig5} (b), the theoretical preditions for $\Delta H$ have
the expected ``Curie tail'' $\Delta H \sim {\cal A}(x)/T$ at low temperature. For uncorrelated site disorder,
the coefficient ${\cal A}$ is expected to scale as $x^2$ for asymptotically small $x$; however, for
not-too-small $x$ relevant to experiment, we find that ${\cal A}(x)$ can be fit 
well by an approximately {\em linear} $x$ dependence ${\cal A}(x) \sim x$.

Thus, our theory reproduces the broad, apparently symmetric lines seen in experiments, with
broadening scaling as $\Delta H \sim x/T$ for not-too-low $x$ and $T$.
Although these experimental facts are fully reproduced, we caution that our
results do not provide a fully quantitative explanation of the NMR data: As is
clear from Fig~\ref{Fig5} (b), our minimal disorder model consisting of
independent vacancies of the nominal experimental density significantly
underestimates the absolute scale of the linewidth $\Delta H$. Since isolated
vacancies lead to no oscillating spin textures scaling as $1/T$\cite{sup}, this discrepancy clearly demonstrates
that the density of orphan spins in the experimental samples 
is significantly higher than that expected from uncorrelated Ga substitution of
the Cr lattice. 
The most frugal resolution is perhaps
that the Ga substitution in the Cr kagome layer may be
correlated. Additionally, and perhaps more realistically, other sources of
disorder, for instance random strain-induced bond randomness, could also affect
the effective density of `orphan spins'. Unfortunately, it does not seem
possible to use 
measurements of the impurity Curie constant $C_d$\cite{schifferSUS} to shed
light on this, as uniform susceptibility measurements  also pick up, apart
from the signal due to Kagome layer orphan spin textures,  a much larger ``background'' contribution from free Cr spins created when Ga impurities substitute
for Cr in the isolated dimer layer (Fig~\ref{Fig1_lattice}) of the SCGO lattice.

In conclusion, we note that these disorder induced orphan-texture complexes
in SCGO embody several central themes of modern condensed matter physics:  The
emergence of new types of extended degrees of freedom, and their rich physics
at 
finite density, is a common strand that runs through diverse  
examples\cite{Physics_today} such as Skyrmion lattices in  
spinor condensates\cite{cherngTHESIS}, itinerant \cite{binzSkyr} and  quantum
Hall  
magnets\cite{Skyrmion_review,Gervais_etal_prl05}, and
spin textures in topological phases of matter\cite{Hasan_topological}.
Moreover, the fact that our textures give rise to a relatively strong magnetic
response---their staggered $1/r^{d-1}$ magnetisation profile  is {\em
  parametrically} stronger than the intrinsic $1/r^d$ spin correlations of the
parent spin-liquid---shows that, like in 
the case of defects evidencing the d-wave nature of the order parameters in the cuprates, imperfections are among the best and most direct probes of exotic correlated states of matter.

The authors thank P.~Mendels for providing the
original NMR data (Fig~\ref{Fig5} b) and  gratefully acknowledge useful discussions with
F.~Bert, D.~Dhar, C.~Henley, P.~Mendels, P.~Schiffer, and A.~Zorko, funding from DST SR/S2/RJN-25/2006,
financial support for collaborative visits from the Fell Fund (Oxford),
the ICTS TIFR (Mumbai), ARCUS (Orsay) and MPIPKS (Dresden), as well as
computational resources at TIFR.

\end{document}